\documentclass[final,5pt,times]{elsarticle}
\usepackage{txfonts}
\usepackage{graphicx}

\usepackage{amssymb}

\begin{document}

\begin{frontmatter}



\title{ Security proof of Counterfactual Quantum Cryptography against General Intercept-resend Attacks and Its Vulnerability}


\author{Sheng Zhang\fnref{fn1}}
\author{Jian Wang}
\author{Chao-Jing Tang}
\author{Quan Zhang}
\address{School of Electronic Science and Engineering, National University of Defense Technology, Changsha 410073, China}
\fntext[fn1]{Tel. No.: 86 0731 84575706, E-mail address: shengzhcn@gmail.com.}

\begin{abstract}
Counterfactual quantum cryptography (CQC), recently proposed by Noh, is featured with no transmission of signal particles. This exhibits evident security advantage, such as its immunity to the well known PNS attack. In this paper, the theoretical security of CQC protocol against the general intercept-resend attacks is proved by bounding the information of an eavesdropper Eve more tightly than in Yin's proposal[Phys. Rev. A 82, 042335 (2010)]. It is also showed that practical CQC implementations  may be vulnerable when equipped with imperfect apparatuses, by proving that a negative key rate can be achieved when Eve launches a time-shift attack based on imperfect detector efficiency.
\end{abstract}


\end{frontmatter}


\section{\label{sec:level1}Introduction}

Quantum key distribution (QKD) \cite{BB84,Ekert91,Gisin02}, which is the most prominent application in quantum information theory, enables two distant parties, conventionally referring to Alice and Bob, to establish a secret key guaranteed by fundamental quantum mechanics, such as the no-cloning theorem. Since the first QKD scheme was proposed in 1984 \cite{BB84}, a lot of attention to QKD has been payed both in theoretical and experimental areas in these decades\cite{Gottesman04,Stucki02,yz06,yz06p,CE03,Poppe08,FXX08}. In Most of the schemes, the information bits are conventionally encoded into the quantum states, e.g. chosen from two conjugated bases, and then transmitted to Bob in the public channel controlled by a powerful eavesdropper Eve. Recently Noh proposed a novel QKD protocol(Noh09 protocol) \cite{Noh09} using a striking phenomenon commonly termed as quantum counterfactual effect, which is initiated by the idea of interaction-free measurement in quantum computation \cite{EV93,Vaidman07}. It is exciting that the information particles in his protocol are not transmitted via any physical channel, thus, the security is witnessed straightforwardly.

In Noh09 protocol, Alice prepares a single photon pulse from two orthogonal states, i.e., a horizontally polarized state $|H>$ and a vertically polarized state $|V>$. The pulse is then split into two by a beam splitter. The initial quantum states after the BS is written by $|\phi_{0(1)}>=\sqrt{T}|0>_{a}|H(V)>_{b}+i\sqrt{R}|H(V)>_{a}|0>_{b}$,  where $|0>_{a(b)}$ denotes the vacuum state in mode $a(b)$. In Bob's secure zone, a polarization beam splitter (PBS) and optical switch (SW) are applied to block path $b$ according to Bob's randomly chosen polarization. Specifically, the sub-pulse in path $b$ is blocked only if their polarizations are identical, otherwise it will be reflected back to Alice. Ideally, there are three intrinsic events: ($E_{1}$) Detector $D_{1}$ clicks, it occurs when the single photon is reflected by the BS and their polarizations are identical; ($E_{2}$) Detector $D_{2}$ clicks, it occurs when Bob's polarization is inconsistent with Alice's, or when Bob's polarization is identical to Alice's and the photon travels in path $a$; ($E_{3}$) Detector $D_{3}$ clicks, it occurs when Bob's polarization is identical to Alice's and the photon travels in path $b$. The raw key is generated from part of the instances in event $E_{1}$. Although a simple security analysis against a special intercept-resent attack was presented in Noh09 protocol,  a more strict one is expected.

Recently, Zhen-qiang Yin \cite{Yin10} et al. gave a security proof of the counterfactual quantum cryptography with the same technique proposed in ref.\cite{Shor00}. In Yin's proposal, the security of Noh09 protocol relies on a so-called equivalent EDP (entanglement distillation protocol), which is designed to simulate the function of a CQC protocol. Let us make a simple review on this EDP. Alice first prepared N pairs of entanglement states written by $|\Psi>_{A}=\frac{1}{\sqrt{2}}[|H>(\sqrt{T}|0>_{a}|H>_{b}+i\sqrt{R}|H>_{a}|0>_{b})+
|V>(\sqrt{T}|0>_{a}|V>_{b}+i\sqrt{R}|V>_{a}|0>_{b})]$, she reserved half of the particles of the entanglement states in her own and transmitted the other half to Bob, after the photons pass through the BS. Bob's blocking operation in Noh09 protocol is abstracted by a unitary operation which is performed on the pulse in path $b$. Another unitary is performed on the entanglement state to check the consistency, the sifted key is generated on the subset of the entanglement states for which the consistency holds. Then its security under the collective attacks was proved in the rest of the paper, by bounding the phase error rate of the quantum states.

We find that there are some flaws in Yin's proposal with two aspects. First, Eve may not try to entangle her ancilla with the intercepted states, since she has no access into mode $a$, thus she is still unable to get any information about the key even if she succeeds in doing this. Note that it is not necessary to entangle her ancilla with the intercepted one for the sake of getting more information. Second,the so-called equivalence between their EDP and the original protocol needs further consideration, ambiguous evidence shows that the outputs from the two protocols differ sometimes. In other words, it is expected to give a equivalence proof from the views of Alice, Bob and Eve.

In this paper, we give a more intuitive security proof of CQC against the general intercept-resend attacks. The paper is organized as follow: In section \ref{attack}, the general intercept-resend attack is introduced, then an attack model is given. Based on this model, the key rate for the ideal protocol is calculated in section \ref{ideal_sec}. In section \ref{prac_sec}, we investigate the practical security of a real-world CQC implementation in a scenario where a time-shift attack based on imperfect detectors is performed. At last, a conclusion is drawn.

\section{Modeling the Attacks}\label{attack}
In a conventional QKD scheme, Eve usually performs a unitary operation on the intecepted qubits and her ancilla, she will not measure her probes until all the classical information is revealed . Thus Eve's attack can be formalized by a unitary operator and an optimal measurement operator conditioned on the classical information. However, this strategy may not account for a CQC protocol, because information carriers are never transmitted via the channel, Eve will always fail to entangle her probe with the right qubits. A more intuitive way is to launch a general intercept-resend attack, with which Eve may skillfully misleed both Alice and Bob as possible as she can to agree with a key on those particles travelling in path $b$, and corrupt as much as possible the instances in event $E_{1}$.

Quantum system is an abstract Hilbert space which includes both actual and fictive subsystems. We denote the initial quantum state in this abstract space by $H_{A}\otimes H_{B}\otimes H_{E}\otimes H_{D}$, where $H_{A}$, $H_{B}$ and $H_{E}$ are the subspaces owned by Alice, Bob and Eve respectively, and $H_{D}$ is a fictive system which describes the measurement results of detectors $D_{1}$, $D_{2}$ and $D_{3}$. Then the initial state can be expressed by

\begin{equation} \label{eq:1}
\begin{array}{rl}
|\Phi_{0}>= & \{|\Xi_{xyz}>(i\sqrt{R}|p>_{a}|0>_{b}+\\
& \sqrt{T}|0>_{a}|p>_{b})|q>_{B}|e>_{E}\}^{\otimes n}.
\end{array}
 \end{equation}
Here $|\Xi_{xyz}>$ is a fictive state where $x$,$y$ and $z$ denote the measurement results of detector $D_{1}$, $D_{2}$ and $D_{3}$ respectively, $p$ and $q$ are variables set in $H$ or $V$, and the variable $e$ denoting Eve's result of measurement takes the value from $\{0,H,V\}$.  Let $n=1$, for simplicity, so that only one-bit  key is analyzed, hence EQ.(\ref{eq:1}) is changed to
\begin{equation} \label{eq:2}
\begin{array}{rl}
|\Phi_{0}>= &|\Xi_{xyz}>(i\sqrt{R}|p>_{a}|0>_{b}+\\
& \sqrt{T}|0>_{a}|p>_{b})|q>_{B}|e>_{E}.
\end{array}
 \end{equation}

Before Eve's attack is modeled, it is necessary to define a unitary operator describing the behaviors of the original protocol,
\begin{equation} \label{eq:3}
\begin{array}{rl}
|\Phi&_{1}>\\
 =& U_{CQC}|\Xi_{xyz}>(i\sqrt{R}|p>_{a}|0>_{b}+\sqrt{T}|0>_{a}|p>_{b})|q>_{B}|e>_{E}\\
 =& \frac{1}{2}\{\sum_{\{p,q|q=p\}}i\sqrt{R}|p>_{a}|0>_{b}|q>_{B}(\sqrt{T}|\Xi_{p00}>+\sqrt{R}|\Xi_{0p0}>)\\
 & +\sum_{\{p,q|q=p\}}\sqrt{T}|0>_{a}|p>_{b}|q>_{B}|\Xi_{00p}>\\
 & +\sum_{\{p,q|q\neq p\}}(i\sqrt{R}|p>_{a}|0>_{b}\\
 & +\sqrt{T}|0>_{a}|p>_{b})|q>_{B}|\Xi_{0p0}>\}\otimes |e>_{E}.
\end{array}
 \end{equation}
Eve's attack is now written by another unitary operator $U_{E}$ acting on the Hilbert space $H_{A}\otimes H_{B}\otimes H_{E}\otimes H_{D}$,
\begin{equation} \label{eq:4}
\begin{array}{rl}
|\Phi&_{out}>\\
 =&U_{E}|\Phi_{1}>\\
 =& \frac{1}{2}\sum_{\{p,q|q=p\}}i\sqrt{RT}|p>_{a}|0>_{b}|q>_{B}(\sum_{\{x,y,z|x=p,y \cup z\neq 0\}}\alpha^{1}_{xyz}|\Xi_{xyz}>\\
 & |0>_{E}+\alpha^{1}_{p00}|\Xi_{p00}>|0>_{E})\\
 & +\frac{1}{2}\sum_{\{p,q|q=p\}}(-R)|p>_{a}|0>_{b}|q>_{B}(\sum_{\{x,y,z|y=p,x \cup z\neq 0\}}\alpha^{2}_{xyz}|\Xi_{xyz}>\\
 & |0>_{E}+\alpha^{2}_{0p0}|\Xi_{0p0}>|0>_{E})\\
 & +\frac{1}{2}\sum_{\{p,q|q\neq p\}}(i\sqrt{R}|p>_{a}|0>_{b}(\sum_{\{x,y,z|x \cup y \neq 0,z\neq 0\}}\alpha^{3}_{xyz}|\Xi_{xyz}>|0>_{E}\\
 &+\alpha^{3}_{p00}|\Xi_{p00}>|0>_{E}+\alpha^{3}_{0p0}|\Xi_{0p0}>|0>_{E})\\
 & +\frac{1}{2}\sum_{\{p,q|q=p\}}\sqrt{T}|0>_{a}|p>_{b})|q>_{B}(\alpha^{4}_{p00}|\Xi_{p00}>|p>_{E}\\
 & +\alpha^{4}_{0p0}|\Xi_{0p0}>|p>_{E}+\alpha^{4}_{00p}|\Xi_{00p}>|p>_{E})\\
 & +\frac{1}{2}\sum_{\{p,q|q\neq p\}}\sqrt{T}|0>_{a}|p>_{b})|q>_{B}(\alpha^{5}_{p00}|\Xi_{p00}>|p>_{E}\\
 & +\alpha^{5}_{0p0}|\Xi_{0p0}>|p>_{E}+\alpha^{5}_{00q}|\Xi_{00q}>|p>_{E}).
\end{array}
 \end{equation}

This model is naturally obtained according to the general intercept-resend strategy, and five cases need to be focused on: (C1) Alice's polarization is consistent with Bob's , Mode b is in vacuum and detector $D_{1}$ is supposed to click. In this case, Eve's attack renders at least two detectors click at the same time, this explains the term $\sum_{\{x,y,z|x=p,y \cup z\neq 0\}}\alpha^{1}_{xyz}|\Xi_{xyz}>|0>_{E}$, the other term $\alpha^{1}_{p00}|\Xi_{p00}>|0>_{E}$ represents the probability of those uncorrupted events. In particular, $|0>_{E}$ denotes the state when Eve gets nothing about the key. (C2) Alice's polarization is consistent with Bob's, mode $b$ is in vacuum and detector $D_{2}$ is supposed to click. Eve's attack is then presented in the term $\sum_{\{x,y,z|y=p,x \cup z\neq 0\}}\alpha^{2}_{xyz}|\Xi_{xyz}>|0>_{E}$. (C3) Alice's polarization is inconsistent with Bob's, mode $b$ is in vacuum and detector $D_{2}$ is supposed to click. similarly, $\sum_{\{x,y,z|x \cup y \neq 0,z\neq 0\}}\alpha^{3}_{xyz}|\Xi_{xyz}>|0>_{E}$ describes the events in which at least two detectors click simultaneously, $\alpha^{3}_{p00}|\Xi_{p00}>|0>_{E}$ or $\alpha^{3}_{0p0}|\Xi_{0p0}>|0>_{E}$ tells that detector $D_{1}$ or $D_{2}$ clicks. (C4) Alice's polarization is consistent with Bob's, mode $b$ is non-vacuum and detector $D_{3}$ is supposed to click. In this case, it is possible for Eve to compromise the key, as the term $\alpha^{4}_{p00}|\Xi_{p00}>|p>_{E}$ says, the explanation to $\alpha^{4}_{00p}|\Xi_{00p}>|p>_{E}$ is that Eve needs to probe Bob's polarization sometimes to adjust her following interceptions. (C5) Alice's polarization is inconsistent with Bob's, mode $b$ is non-vacuum and detector D2 is supposed to click. It is similar with the case C4 except that Eve is unable to get any information from her results in this case.

We must point out that this formalization is general, or is valid for any intercept-resend attack in other words, because the coefficients  $\{\alpha^{i}_{xyz}|i=1,2,3,4,5\}$ are arbitrary. In the next section, we will bound Eve's information based on this model, and give a more rigorous key rate of CQC under the condition that all the quantum apparatuses are perfect.

\section{Theoretical security against general intercept-resend attacks}\label{ideal_sec}

In a security proof of conventional quantum key distribution schemes, such as BB84 protocol, security definition is important and an asymptotic security is ultimately concluded as a function of the length of the quantum sequence. In this paper, the infinite-length security rather than an asymptotic one is investigated.

Suppose all the quantum apparatuses are perfect, especially the detectors, of which the efficiency $\eta$ reaches 1 and the dark counter rate $p_{d}$ is set to zero, any event in which more than one detectors click is definitely absent, because it implies the existence of an eavesdropper Eve. Thereby the  model is simplified after fixing the terms as $\{\alpha^{1}_{xyz}=0|x=p,y \cup z\neq 0\}$, $\{\alpha^{2}_{xyz}=0|y=p,x \cup z\neq 0\}$ and $\{\alpha^{3}_{xyz}|z \cup y\neq 0, z\neq 0\}$. Then one obtains
\begin{equation} \label{eq:5}
\begin{array}{rl}
|\Phi&_{out}>\\
 =&U_{E}|\Phi_{1}>\\
 =& \frac{1}{2}\sum_{\{p,q|q=p\}}i\sqrt{RT}|p>_{a}|0>_{b}|q>_{B}\alpha^{1}_{p00}|\Xi_{p00}>|0>_{E}\\
 & +\frac{1}{2}\sum_{\{p,q|q=p\}}(-R)|p>_{a}|0>_{b}|q>_{B}\alpha^{2}_{0p0}|\Xi_{0p0}>|0>_{E}\\
 & +\frac{1}{2}\sum_{\{p,q|q\neq p\}}(i\sqrt{R}|p>_{a}|0>_{b}(\alpha^{3}_{p00}|\Xi_{p00}>|0>_{E}\\
 & +\alpha^{3}_{0p0}|\Xi_{0p0}>|0>_{E})\\
 & +\frac{1}{2}\sum_{\{p,q|q=p\}}\sqrt{T}|0>_{a}|p>_{b})|q>_{B}(\alpha^{4}_{p00}|\Xi_{p00}>|p>_{E}\\
 & +\alpha^{4}_{0p0}|\Xi_{0p0}>|p>_{E}+\alpha^{4}_{00p}|\Xi_{00p}>|p>_{E})\\
 & +\frac{1}{2}\sum_{\{p,q|q\neq p\}}\sqrt{T}|0>_{a}|p>_{b})|q>_{B}(\alpha^{5}_{p00}|\Xi_{p00}>|p>_{E}\\
 & +\alpha^{5}_{0p0}|\Xi_{0p0}>|p>_{E}+\alpha^{5}_{00q}|\Xi_{00q}>|p>_{E}).
\end{array}
 \end{equation}
This model is correct, for instance in the ideal case where there is no eavesdropper one may obtain
\begin{equation} \label{eq:6}
\begin{array}{rl}
p_{D_{1}}=\frac{RT}{2},
\end{array}
 \end{equation}
 \begin{equation} \label{eq:7}
\begin{array}{rl}
p_{D_{2}}=\frac{R^{2}}{2}+\frac{1}{2},
\end{array}
 \end{equation}
\begin{equation} \label{eq:8}
\begin{array}{rl}
p_{D_{3}}=\frac{T}{2},
\end{array}
 \end{equation}
\begin{equation} \label{eq:9}
\begin{array}{rl}
p^{1}_{e}=p^{2}_{e}=p^{3}_{e}=0.
\end{array}
 \end{equation}

Note that any pair of inconsistent measurement results of Alice and Bob indicates an error bit, it is easy to confirm six important parameters which are crucial to bound Eve's information $I_{AE}$,
\begin{equation} \label{eq:10}
\begin{array}{rl}
p_{D_{1}} =\frac{RT}{2}|a^{1}_{p00}|^{2}+\frac{R}{2}|a^{3}_{p00}|^{2}+\frac{T}{2}|a^{4}_{p00}|^{2}+\frac{T}{2}|a^{5}_{p00}|^{2},
\end{array}
 \end{equation}
\begin{equation} \label{eq:11}
\begin{array}{rl}
p_{D_{2}} =\frac{R^{2}}{2}|a^{2}_{0p0}|^{2}+\frac{R}{2}|a^{3}_{0p0}|^{2}+\frac{T}{2}|a^{4}_{0p0}|^{2}+\frac{T}{2}|a^{5}_{0p0}|^{2},
\end{array}
 \end{equation}
\begin{equation} \label{eq:12}
\begin{array}{rl}
p_{D_{3}} =\frac{T}{2}(|a^{4}_{00p}|^{2}+a^{5}_{00q}|^{2}),
\end{array}
 \end{equation}
\begin{equation} \label{eq:13}
\begin{array}{rl}
p^{1}_{e}=\frac{R}{2}|a^{3}_{p00}|^{2}+\frac{T}{2}|a^{5}_{p00}|^{2},
\end{array}
 \end{equation}
 \begin{equation} \label{eq:14}
\begin{array}{rl}
p^{2}_{e}=0,
\end{array}
 \end{equation}
 \begin{equation} \label{eq:15}
p^{3}_{e}=\frac{T}{2}|a^{5}_{00q}|^{2}.
 \end{equation}

Now the key rate, as well as the bound of Eve's information, can be given by the following theorem,

\textbf{Theorem 1} For a given observable probability distribution \{$P_{D_{1}}$ , $P_{D_{2}}$ , $P_{D_{3}}$\}  and the error rates $P^{1}_{e}$  ,$ P^{2} _{e}$ and $P^{3}_{e}$, the mutual information between Alice and Bob is $I_{AB}=p_{D_{1}}^{\ast }[1-h(\frac{p^{1}_{e}}{p_{D_{1}}})]$, and between Alice and Eve is $I_{AE}=(\frac{T}{2}+p^{3}_{e}-p_{D_{3}})R$, hence the infinite-effect key rate is

\begin{equation} \label{eq:16}
m_{k}=I_{AB}-I_{AE}=p_{D_{1}}-(\frac{T}{2}+p^{3}_{e}-p_{D_{3}})R-p_{D_{1}}^{\ast }h(\frac{p^{1}_{e}}{p_{D_{1}}}).
 \end{equation}

Proof: See \ref{app1}.

\begin{figure}
  \centerline{\includegraphics{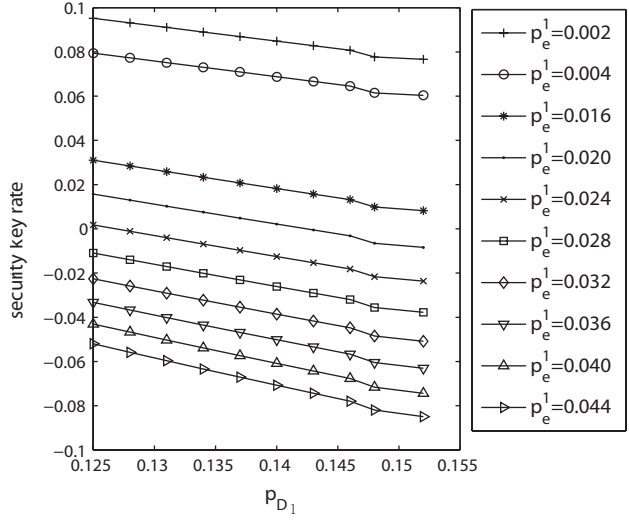}}
  \caption{Theoretical key rate of CQC as a function of $P_{D_{1}}$}\label{IKR}
\end{figure}

So far we have proved the security of CQC against the general intercept-resend attacks in the context that quantum apparatus works according to its specification. As we can see in Eq. (\ref{eq:16}), the key rate is inherently bounded by the probabilities of events $E_{1}$, $E_{2}$ and $E_{3}$, and the error rates in event $E_{1}$ and $E_{3}$ ( Eve's attack does not affect event $E_{2}$, seeing from Eq.(\ref{eq:14})). This coincides with the real situation in the sense that any abnormal probability distribution or a relatively high error rate in one of events immediately implies the existence of an eavesdropper. Also note that any protocol is secure only if it satisfies $m_{k}>0$. Fortunately, as shown in Fig.\ref{IKR}, a positive security key rate can be achieved when the error rates in event $E_{1}$ and $E_{3}$ are small enough and the probability distribution is not far from normal, i.e., $p_{D_{1}}=RT/2$, $p_{D_{2}}=R^{2}/2+1/2$ and $P_{D_{3}}=T/2$.

\section{A time-shift attack in practical CQC system}\label{prac_sec}
We have been reminded that imperfect detectors can be the loophole in practical QKD systems \cite{Makarov06}, several elaborately conceived attacks can be applied to crack the whole system without the detection \cite{Makarov08,Fung07,Fung07pra}. Unfortunately, such problem also exists in practical CQC implementations. In this section, it will be showed that practical CQC systems are vulnerable to device imperfections by an example of a time-shift attack based on imperfect detector efficiency.

Due to the dark counter rate, two or three detectors might click simultaneously, this event is denoted by $E_{4}$. Assume that Eve is as powerful as only constrained by quantum physics. Naturally, she must lower the probability $P_{E_{4}}$ to make herself undetectable, thus it should satisfy
\begin{equation} \label{pd}
P_{E_{4}}\leqslant 2P_{d},
  \end{equation}
i.e.,
\begin{equation} \label{eq:17}
\begin{array}{rl}
p_{D4}=& \frac{RT}{2}\sum_{\{x,y,z|x=p,y\cup z\neq 0\}}\alpha^{1}_{xyz}+\frac{R^{2}}{2}\sum_{\{x,y,z|y=p,x\cup z\neq 0\}}a^{2}_{xyz}\\
&+\frac{R}{2}\sum_{\{x,y,z|x\cup y\neq 0,z\neq 0\}}a^{3}_{xyz}\leqslant 2P_{d}.
\end{array}
  \end{equation}

Event $E_{4}$ is advantageous to Eve, yet as being bounded by dark counter rate, the corrupted information seems trivial, otherwise Eve takes high risk being detected. Lemma 1.1 gives the maximal corrupted information by Eve.\\
\textbf{Lemma 1.1} For a given dark counter rate $p_{d}$ of the detector, the maximal corrupted bit rate is obtained by \\
\begin{equation} \label{lemma1}
r_{c max}=(T+1)P_{d}+[p_{D_{1}}-(T+1)P_{d}] h(\frac{p^{1}_{e}}{p_{D_{1}}-(T+1)P_{d}})-p_{D_{1}} h(\frac{p^{1}_{e}}{p_{D_{1}}}).
  \end{equation}
Proof: See \ref{app2}.

\begin{figure}
  \centerline{\includegraphics{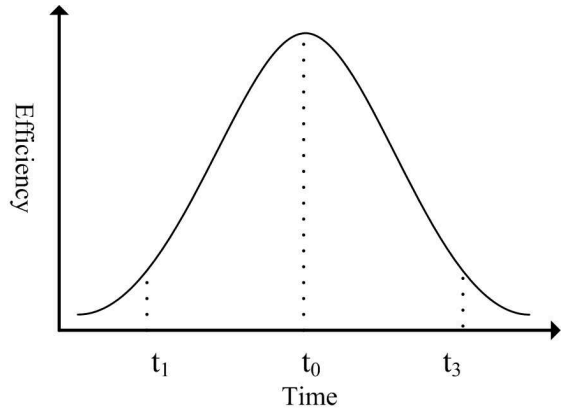}}
  \caption{Efficiency of an imperfect detector as a function of time.}\label{EFFCY}
\end{figure}

Another important parameter of the detector is the efficiency $ \eta$, which brings the risk: Eve may artificially manipulate her device to lower the efficiency $\etaup$, in order to pass the test with a greater probability. To compromise the key, Eve has to probe Bob's polarization by sending fake photons, then these photons are unlikely to be caught by detector $D_{3}$ if the efficiency of detector $D_{3}$ is low enough. Therefor Eve is able to control an imperfect detector simply by using an optical delay, i.e., she may launch a time-shift attack \cite{Fung07}to artificially lower the efficiency to an acceptable level, due to the distinguishability between the real efficiency function (refer to Fig.\ref{EFFCY}) and the ideal one. This evidently causes an increment on the mutual information between Alice and Eve, demonstrated by Lemma 1.2.\\
\textbf{Lemma 1.2} Given a prior fixed detector efficiency $\etaup$ , the increment of the mutual information between Alice and Eve is
\begin{equation} \label{eq:18}
\bigtriangleup I^{\eta}_{AE}=\frac{1-\eta}{\eta}(p_{D_{3}}-p^{3}_{e})R.
 \end{equation}
Proof: See \ref{app3}.\\

Combined with Lemma 1.1 and Lemma 1.2£¬the total of the reduced information $m_{k}$  is immediately obtained by a sum of the two,
\begin{equation} \label{eq:19}
\bigtriangleup m_{k}=\gamma_{c max}+\bigtriangleup I^{\eta}_{AE}.
 \end{equation}
So the key rate under a time-shift attack is changed to
\begin{equation} \label{eq:20}
m^{'}_{k}=m_{k}-\bigtriangleup m_{k}.
 \end{equation}
\begin{figure}
  \centerline{\includegraphics{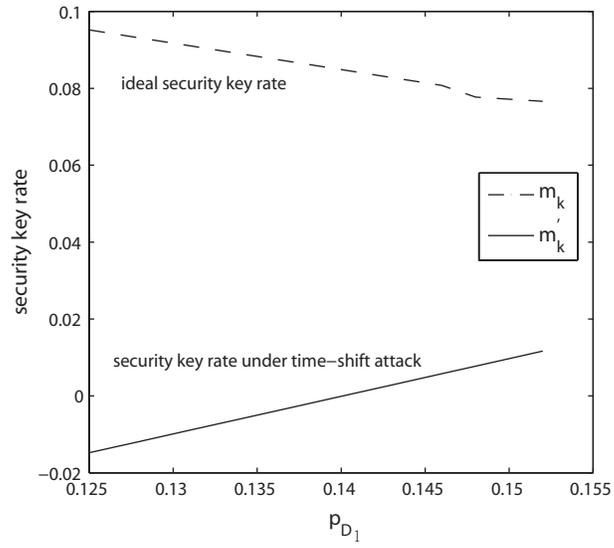}}
  \caption{The key rate under a time-shift attack.}\label{RKR}
\end{figure}

\begin{figure}
  \centerline{\includegraphics{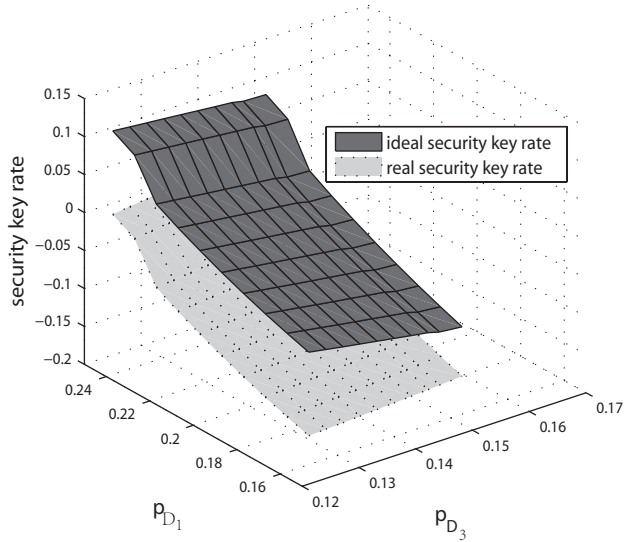}}
  \caption{security key rate under time-shift attack as functions of $P_{D_{1}}$  and  $P_{D_{3}}$, comparing to the ideal one. }\label{Comp}
\end{figure}

As shown in Fig.\ref{RKR}, a negative security key rate can be achieved. Consequently, Eve can obtain full information about the key even if $m_{k}>0$. A 3-D illustration of the key rate as functions of $P_{D_{1}}$  and  $P_{D_{3}}$ is showed in Fig.\ref{Comp} for a better understanding of this conclusion. Since there may be , for most of the QKD protocols, unexpected difficulties when we try to fill the gap between the ideal model and a realistic implementation to a negligible level at present, it is expected that more applicable CQC protocols are required in the future.\\

\section{Conclusion}
We have proved the security of counterfactual quantum cryptography against general intercept-resend attacks, by modeling it with two unitary operations, which are abstracted from Noh09 protocol and the attack strategy respectively. Note that collective attacks in conventional QKD protocols do not account for CQC equally, yet intercept-resend attack seems to be more advantageous to the eavesdropper. Intuitively, Eve may corrupt the key-to-be events and mislead the key generation. The protocol is proved to be secure against such attacks in the context that quantum apparatuses are perfect, then the security key rate is bounded by the probability distribution and the error rates of three intrinsic events. Our result is complementary to the one in Ref.\cite{Yin10}, in which the error rate in events $D_{3}$ is not involved in the conclusion. Also note that Eve's attack not only alters the probability distribution, but also introduces errors, thus, our proof coincides with the real situation more smoothly. Being demonstrated by the example of a time-shift attack, it is also showed that practical CQC implementation is vulnerable to imperfect quantum apparatuses, this opens another problem how to put forward a secure CQC system in real life.
\section{acknowledgement}
This work is supported by the National Natural Science Fundation of China with the project number 60872052. The authors are grateful to professor Lev Vaidman for informing us about the importance of interaction-free measurement on which CQC is based.
\appendix

\section{\label{app1}}
\textbf{Theorem 1} For a given observable probability distribution \{$P_{D_{1}}$ , $P_{D_{2}}$ , $P_{D_{3}}$\}  and the error rates $P^{1}_{e}$  ,$ P^{2} _{e}$ and $P^{3}_{e}$, the mutual information between Alice and Bob is $I_{AB}=p_{D_{1}}^{\ast }[1-h(\frac{p^{1}_{e}}{p_{D_{1}}})]$, and between Alice and Eve is $I_{AE}=(\frac{T}{2}+p^{3}_{e}-p_{D_{3}})R$, hence the infinite-effect key rate is
\begin{equation} \label{eq:21}
m_{k}=I_{AB}-I_{AE}=P_{D_{1}}[1-h(\frac{P^{1}_{e}}{P_{D_{1}}})].
 \end{equation}
\emph{Proof}. To proof Theorem 1, we start with Eq. (\ref{eq:5}) and fix an important parameter $\alpha^{4}_{\rho00}$ , which is given by the unitarity  of  $U_{E}$ shown below\\
\begin{equation} \label{eq:22}
|\alpha^{4}_{p00}|^{2}+|\alpha^{4}_{0p0}|^{2}+|\alpha^{4}_{00p}|=1.
 \end{equation}
According to case (C4), which is the only one Eve can be successful in, when the fake photon transmitted by Eve is received by Alice, the probabilities for the two probabilities are immediately set by the BS,
\begin{equation} \label{eq:23}
|\alpha^{4}_{p00}|^{2}/|\alpha^{4}_{0p0}|^{2}/=R/T,
 \end{equation}
combine with Eq. (\ref{eq:13}), (\ref{eq:15}), (\ref{eq:17}) and (\ref{eq:18}), we obtain
\begin{equation} \label{eq:24}
|\alpha^{4}_{p00}|^{2}=(\frac{T}{2}+P^{3}_{e}-P_{D_{3}})R.
 \end{equation}
With the binary shannon  entropy function, Eve's information  $I_{AE}$ is therefor bounded by
 \begin{equation} \label{eq:25}
\begin{array}{rl}
I_{AE}=&I(p:e|D_{1}=1,C4=1)\\
 =& p(E_{1}=1,C4=1)[1-h(p(\bar{e}|p)]\\
 =& p(E_{1}=1,C4=1)   ,\\
 =&\frac{T}{2}|\alpha|^{4}_{p00}\\
 =& (\frac{T}{2})+P^{3}_{e}-P_{D_{3}}R,
\end{array}
 \end{equation}
where $E_{1}=1$ denote that detector $D_{1}$ clicks , $C4=1$ means that Eve confirms that case C4 is occupied, and $h(x)=-x\log x-(1-x)\log(1-x)$ is the binary Shannon entropy function. Obviously, Eve obtains no information when $p_{D_{3}}$ is normal and there is no error in event $E_{3}$. Thereby it is natural to obtain that $I_{AE}=0$  if $P_{D_{3}}=\frac{T}{2}$ and $P^{3}_{e}=0$ .\\

To bound the mutual information between Alice and Bob, i.e., $I_{AB}$ , let us first find the
QBER defined by $QBER=\frac{p(error)}{p(arrive)}$, which is different with the one in conventional QKD schemes. Since only event $E_{1}$ is useful for the key generation, the QBER can be equalized to the error rate for event $E_{1}$, thus, the actual QBER for CQC protocol is obtained by\\

 \begin{equation} \label{eq:26}
QBER_{D_{1}=1}=p(error|D_{1}=1)=\frac{P^{1}_{e}}{P_{D_{1}}}.
 \end{equation}

With the similar technique of bounding Eve's information in APPENDIX \ref{app1}, we can obtain $I_{AB}$

 \begin{equation} \label{eq:27}
I_{AB}I(q:p|D_{1}=1)=p_{D_{1}}[1-h(QBER_{D_{1}=1})],
 \end{equation}

where  $QBER_{D_{1}=1}$ is given by Eq.(\ref{eq:26}).

In particular, we could, from Eq.(\ref{eq:27}), obtain $I_{AB}=\frac{RT}{2}$ given $QBER_{D1=1}=0$. This is consistent with the result for the ideal case. Now the key rate is naturally obtained by

 \begin{equation} \label{eq:28}
 \begin{array}{rl}
 m_{k}=&I_{AB}-I_{AE}\\
 =&P_{D_{1}}-(\frac{T}{2}+P^{3}_{e}-P_{D_{3}})R-P_{D_{1}}h(\frac{P^{1}_{e}}{P}_{D_{1}}).
 \end{array}
 \end{equation}

 This concludes the proof.

\section{\label{app2}}

\textbf{Lemma 1.1} For a given dark counter rate $p_{d}$ of the detector, the maximal corrupted bit rate is obtained by
\begin{equation} \label{eq:29}
\begin{array}{rl}
\gamma_{c max}=&(T+1)P_{d}+[P_{D_{1}}-(T+1)P_{d}]\\
& \ast h(\frac{P^{1}_{e}}{P_{D_{1}}-(T+1)P_{d}})-P_{D1}h(\frac{P^{1}_{e}}{P_{D_{1}}}).
\end{array}
 \end{equation}

\emph{Proof}. Before we start the proof, a discussion about Eq.(\ref{eq:4}) is presented. It is shown that Event $E_{4}$ occurs only when mode $b$ is in vacuum, and Eve cannot distinguish Cases C1 , C2 and C3, her corruption to the key-to-be events in case C1 absolutely affects the other two. It is not difficult to find that the effect to the three cases is supposed to be the same as a result of Eve's corruption, then we obtain\\

\begin{equation} \label{eq:30}
\begin{array}{rl}
\sum_{\{x,y,z|x=p,y\bigcup z\neq 0\}}|\alpha^{1}_{xyz}|^{2}
 =& \sum_{\{x,y,z|y=p,x\bigcup z\neq 0\}}|\alpha^{2}_{xyz}|^{2}\\
 =& \sum_{\{x,y,z|x\bigcup y\neq 0,z\neq 0\}}|\alpha^{3}_{xyz}|^{2}.
 \end{array}
 \end{equation}

To reach the maximum, the total probability for event $E_{4}$ is set to be

\begin{equation} \label{eq:31}
P_{E_{4}}=2P_{d}.
 \end{equation}

From Eq.(\ref{eq:4}),(\ref{eq:30}) and (\ref{eq:31}), the following expression can be confirmed

\begin{equation} \label{eq:32}
\begin{array}{rl}
\sum_{\{x,y,z|x=p,y\bigcup z\neq 0\}}|\alpha^{1}_{xyz}|^{2}
 =& \sum_{\{x,y,z|y=p,x\bigcup z\neq 0\}}|\alpha^{2}_{xyz}|^{2}\\
 =& \sum_{\{x,y,z|x\bigcup y\neq 0,z\neq 0\}}|\alpha^{3}_{xyz}|^{2}\\
 =&\frac{2}{R}P_{d}.
 \end{array}
 \end{equation}

In addition, from theorem 1 the mutual information between Alice and Bob when the corruption occurs is\\
\begin{equation} \label{eq:33}
I^{real}_{AB}=P^{real}_{D_{1}}[1-h(\frac{P^{1}_{ereal}}{P^{real}_{D_{1}}})],
 \end{equation}
where  $P^{real}_{D_{1}}$ denotes the real probability for event $E_{1}$ and $P^{1}_{ereal}$ is the corresponding error rate. Next, we consider the ideal case, correspondingly, the two parameters are changed to\\
\begin{equation} \label{eq:34}
\begin{array}{rl}
P^{real}_{D_{1}}=&P^{ideal}_{D_{1}}-(\frac{RT}{2}\sum_{\{x,y,z|x=p,y\cup z\neq 0\}}|\alpha^{1}_{xyz}|^{2}\\
& +\frac{R}{2}\sum_{\{x,y,z|x\cup y\neq 0,z\neq 0\}}|\alpha^{3}_{xyz}|^{2}),
\end{array}
 \end{equation}
\begin{equation} \label{peideal}
P^{1}_{eideal}=P^{1}_{ereal}.
 \end{equation}
Explanation to Eq.(\ref{eq:34}) follows: It is assumed the term  $\sum_{\{x,y,z|x\bigcup y\neq 0,z\neq 0\}}|\alpha^{3}_{xyz}|^{2}$ only affects $|\alpha^{3}_{p00}|^{2}$  and cares nothing about $|\alpha^{3}_{0p0}|^{2}$,  since it is advantageous to Eve. Similarly, $\frac{R}{2}\sum_{\{x,y,z|x\cup y\neq 0,z\neq 0\}}|\alpha^{3}_{xyz}|^{2}$ should not affect $|\alpha^{3}_{p00}|^{2}$. Now the mutual information for the ideal case is given by
 \begin{equation} \label{e7}
I^{ideal}_{AB}=P^{ideal}_{D_{1}}[1-h(\frac{P^{1}_{ereal}}{P^{ideal}_{D_{1}}})].
 \end{equation}
Combined with Eq.(\ref{eq:32}),(\ref{eq:33}),(\ref{eq:34}) and (\ref{peideal}), the total corruption rate $\gamma_{c max}$ defined by $\gamma_{c max}=I^{ideal}_{AB}-I^{real}_{AB}$  is bounded, i.e., we have
\begin{equation} \label{e8}
\begin{array}{rl}
\gamma_{c max}=&I^{ideal}_{AB}-I^{real}_{AB}\\
 =&(T+1)P_{d}\\
 & +[P^{ideal}_{D1}-(T+1)P_{d}]h(\frac{P^{1}_{eideal}}{P^{ideal}_{D_{1}}-(T+1)P_{d}})\\
 & -P_{D_{1}}h(\frac{P^{1}_{eideal}}{P^{ideal}_{D_{1}}}).
\end{array}
\end{equation}
Substituting $P^{ideal}_{D_{1}}$ and $P^{1}_{eideal}$  with $P_{D_{1}}$ and  $P^{1}_{e}$ completes the proof.

\section{\label{app3}}
\textbf{Lemma 1.2} Given a prior fixed detector efficiency $\etaup$ , the increment of the mutual information between Alice and Eve is\\
 \begin{equation} \label{1}
\bigtriangleup I^{\eta}_{AE}=\frac{1-\eta}{\eta}(P_{D3}-P^{3}_{e})R.
 \end{equation}
\emph{Proof}. Note that Eve performs a time-shift attack to control the timing of the detector indirectly. In this case, the efficiecny is lowered, thereby the probability for event $E_{3}$ is reduced to
\begin{equation} \label{2}
p_{D3}^{real}=\eta p_{D3}^{ideal}.
 \end{equation}
$P^{3}_{ereal}$  and  $P^{3}_{eideal}$ are defined to denote the probability for the real case and the ideal case respectively.
Similarly, the error rate for event $E_{3}$ is
\begin{equation} \label{3}
p_{ereal}^{3}=\eta p_{eideal}^{3}.
 \end{equation}
From Theorem 1, one immediately obtains
  \begin{equation} \label{4}
I^{ideal}_{AE}=(\frac{T}{2}+P^{3}_{eideal}-P^{ideal}_{D_{3}})R.
 \end{equation}
Similarly, Eve's information in a real case is given by
\begin{equation} \label{5}
I^{real}_{AE}=(\frac{T}{2}+P^{3}_{ereal}-P^{real}_{D3})R.
 \end{equation}
Now combined with Eq.(\ref{2}), (\ref{3}) and (\ref{4}), the increment of the total information of Eve defined by $\bigtriangleup I^{\eta}_{AE}=I^{real}_{AE}-I^{ideal}_{AE}$  is
\begin{equation} \label{6}
\bigtriangleup I^{\eta}_{AE}=\frac{1-\eta}{\eta}(P^{real}_{D_{3}}-P^{3}_{ereal})R.
 \end{equation}
Substituting the labels  $P^{real}_{D_{3}}$ and  $P^{3}_{ereal}$ with $P_{D_{3}}$ and $P^{3}_{e}$ completes the proof.



\bibliographystyle{model1a-num-names}
\bibliography{my_bib_database}







\end{document}